\documentclass[preprint,12pt]{elsarticle}
\biboptions{numbers,sort&compress}

\DeclareGraphicsExtensions{.tif,.png,.jpg,.eps}
\graphicspath{{./Figures/pngs/}}

\usepackage{color,soul}

\usepackage{amssymb}
\usepackage{amsmath}
\usepackage{mathrsfs}

\usepackage{amsthm}

\journal{Journal of Quantitative Spectroscopy and Radiative Transfer}

\begin{document}

\begin{frontmatter}



\title{Near-Field Enhanced Thermionic Energy Conversion for Renewable Energy Recycling}

\author[1]{Mohammad Ghashami}
\author[2]{Sung Kwon Cho}
\author[1]{Keunhan Park}

\address[1]{University of Utah, Department of Mechanical Engineering, Salt Lake City, UT 84112, USA}
\address[2]{University of Pittsburgh, Mechanical Engineering \& Materials Science Department, Pittsburgh, PA 15261, USA}
\ead{kpark@mech.utah.edu} 

\begin{abstract}
This article proposes a new energy harvesting concept that greatly enhances thermionic power generation with high efficiency by exploiting the near-field enhancement of thermal radiation. 
The proposed near-field enhanced thermionic energy conversion (NETEC) system is uniquely configured with a low-bandgap semiconductor cathode separated from a thermal emitter with a subwavelength gap distance, such that a significant amount of electrons can be photoexcited by near-field thermal radiation to contribute to the enhancement of thermionic current density. 
We theoretically demonstrate that the NETEC system can generate electric power at a significantly lower temperature than the standard thermionic generator, and the energy conversion efficiency can exceed 40\%. 
The obtained results reveal that near-field photoexcitation can enhance the thermionic power output by more than 10 times, making this hybrid system attractive for renewable energy recycling.
\end{abstract}


\end{frontmatter}



\section{Introduction}

The current average global energy demand is approximately 14 TW, and is expected to double by 2050~\cite{baxter2009nanoscale}. 
When considering that more than 80\% of energy generation relies on fossil fuels, the unstable prices of these energy sources and increasing carbon dioxide emission pose a grave threat to the global economy and environment. 
It is imperative to develop carbon-free, high-efficiency and low-cost renewable energy harvesting technologies. 
In particular, recycling energy from waste heat may present a great opportunity for energy conservation, as about 20-50\% of energy consumption is being lost to heat in a wide temperature range from lower than 500~K to above 1500~K~\cite{johnson2008waste}.

As a novel approach to directly convert heat to electricity, near-field thermal radiation has been implemented to thermophotovoltaic (TPV) energy conversion~\cite{park2013fundamentals}.
Previous studies have found that thermal radiation can exceed the blackbody limit by several orders of magnitude when objects are separated by a sub-wavelength vacuum gap distance \cite{tien1967effect,loomis1994theory,pendry1999radiative,zhang2007nano,shchegrov2000near,volokitin2004resonant,Joulain2005b,fu2006nanoscale,francoeur2008role}. 
Near-field thermophotovoltaic (NTPV) energy conversion makes use of the near-field enhancement of thermal radiation by placing a low-bandgap photovoltaic cell (or TPV cell) in the near-field of a thermal emitter \cite{laroche2006near, park2008performance, ilic2012overcoming, francoeur2011thermal}.
For example when a In$_{0.18}$Ga$_{0.82}$Sb TPV cell is tens of nanometers away from a tungsten thermal emitter at 2000~K, the TPV cell can generate electric power in the order of 10~W/cm$^2$, which is 20 to 30 times larger than the far-field TPV operating at the same temperature \cite{park2008performance,francoeur2011thermal}. 
The estimated efficiency of the NTPV system is about 20\%, which is better than other solid-state energy conversion devices, such as thermionic ($<\sim$13\%) \cite{lee2009size} and thermoelectric ($<\sim$15\%)~\cite{chen2006nanoscale} generators.
However, the experimental validation of NTPV power generation is not fully convincing yet, mainly due to challenges in realizing a near-field gap distance between a TPV cell and a high-temperature thermal emitter at 1000~K or higher. Although significant progresses have been made in the experimental investigation of near-field thermal radiation between plane structures \cite{ito2015parallel,bernardi2016radiative,ottens2011near,song2016radiative}, experiments have yet to meet the required operating conditions of a NTPV system. For example, one recent work could measure near-field thermal radiation between planar structures having a sub-100nm gap distance, but the measurement was conducted for a 48$\mu$m$\times$48$\mu$m sample area under the temperature difference of 2~K \cite{song2016radiative}. Another challenging issue of the NTPV system is the effective thermal management of the TPV cell when it is in the near-field of a high-temperature thermal emitter. Since a TPV cell is essentially a low-bandgap photovoltaic cell, overheating of the cell by significant heat transfer across the nanoscale gap, including parasitic heat conduction through spacers, may lead to the serious deterioration of its performance. Intensive cooling of the TPV cell is strongly required, often demanding high heat transfer coefficients in the range of $\sim$10$^3$-10$^4$~W/m$^{2}$-K depending on the heat source temperature \cite{francoeur2011thermal,bernardi2015impacts}.

This article presents an alternative way of exploiting near-field thermal radiation for renewable energy harvesting. In particular, we propose near-field enhanced thermionic energy conversion (NETEC) to generate electricity directly from moderate to high temperature heat sources (i.e., 800~K$<T<$1600~K). 
Conventional thermionic energy conversion generates electricity by collecting electrons that are thermally emitted over the potential energy barrier (i.e., work function) of a hot cathode~\cite{hatsopoulos1979thermionic}. 
However, thermionic converters have not been widely implemented due to their general requirement of high cathode temperatures (i.e., $>$1500~K) and a low conversion efficiency below $\sim$15$\%$, even at such high temperatures. 
As a novel technology to circumvent these challenges, photon-enhanced thermionic emission has been recently proposed and received keen attention~\cite{schwede2010photon,segev2012efficiency,segev2013loss,su2013parametric,segev2013high,zhuravlev2014photon}. 
In photon-enhanced thermionic emission, a \textit{p}-type semiconductor cathode is illuminated by high-intensity light to make use of photon energy absorption for thermionic emission. 
Photoexcitation of electrons by above-bandgap photon energy increases the electron concentration in the conduction band. In addition, excess photon energy above the bandgap is converted to heat through thermalization, contributing to the thermal excitation of electrons. The absorption of sub-bandgap photon energy due to impurities or lattice vibrations is another heat source to the cathode. 
Since the broad band of the photon energy absorbed by the cathode is used for thermionic power generation, photon-enhanced thermionic energy conversion has a significantly higher efficiency than conventional thermionic energy conversion: for example, when $\times$1000 concentrated solar radiation is used as a light source, photon-enhanced thermionic emission can yield the efficiencies exceeding 40\% below a cathode temperature of 1000~K \cite{schwede2010photon,segev2012efficiency}.
However, photon-enhanced thermionic emission has not been considered for waste heat recovery to date mainly owing to the low energy density of thermal radiation emitted from a waste heat source. 
In this article, we address this challenge by combining near-field thermal radiation with photon-enhanced thermionic energy conversion, ultimately realizing photo-thermionic energy conversion for renewable energy recycling. The feasibility of the proposed NETEC system is theoretically demonstrated by calculating its power output and energy conversion efficiency. The details of the theoretical background,  modeling and computation results and discussion will be described in the following sections.

\section{Modeling}
\begin{figure}[t!]
\centering
\includegraphics[width=0.75\linewidth]{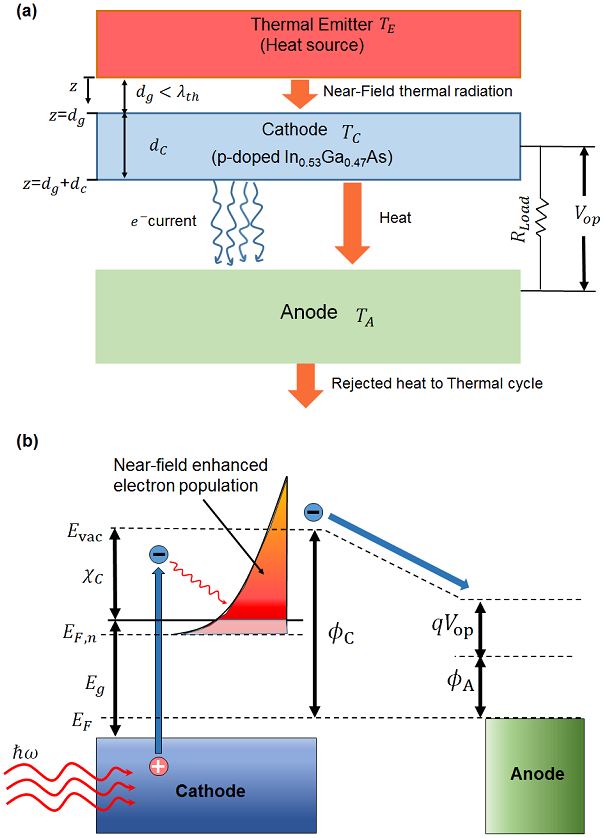}
\caption{(a) Schematics of a NETEC system illustrating its heat transfer and carrier transport mechanisms (not scaled).~(b) Energy diagram showing the work functions and energy barriers for both the cathode and anode. Fermi-level splitting affects the cathode work function and the output voltage since the output voltage equals the difference between cathode and anode work functions.
\label{Fig.1}}
\end{figure}

Figure \ref{Fig.1}(a) illustrates the configuration of the NETEC system, which has a \textit{p}-doped low-bandgap semiconductor cathode separated from a metallic anode with a vacuum gap. The unique feature of the NETEC is a sub-wavelength gap distance between the cathode and a heat source (thermal emitter) to allow near-field radiative energy transfer between them.
We hypothesize that this near-field thermal radiation can photoexcite a significant amount of electrons to the cathode conduction band if the cathode is made of a \textit{p}-doped low-bandgap semiconductor, such as InSb ($E_g =$0.17~eV), GaSb (0.726~eV) or InAs (0.354~eV), where  $E_g$ is the bandgap energy~\cite{levinshtein1996handbook}.
Figure~\ref{Fig.1}(b) demonstrates the energy diagram of the NETEC process. The Fermi level of a \textit{p}-doped semiconductor is lower than the intrinsic level and can be expressed as,
\begin{equation}
E_{F}=E_g/2-k_B T_C\ln(n_p/n_i)
\end{equation}
\noindent where $k_B$ is the Boltzmann constant, $T_C$ is the cathode temperature, $n_p$ is the doping concentration, and $n_i = \sqrt{N_C N_V}\exp\left(-E_g/k_BT_C\right)$ is the intrinsic carrier concentration. Here, $N_C = 2\left({m_n^*k_BT_C}/2\pi{\hbar^2}\right)^{3/2}$ and $N_V = 2\left({m_p^*k_BT_C}/2\pi{\hbar^2}\right)^{3/2}$ are the effective densities of states in the conduction and valence bands, respectively, where $m_n^*$ ($m_p^*$) is the effective mass of electron (hole) carriers and $\hbar = h/2\pi$ is the reduced Planck constant.  
For the present study, we chose In$_{0.53}$Ga$_{0.47}$As doped with beryllium (Be) at $n_p=2.4\times10^{18}$~cm$^{-3}$ as a cathode material due to its low bandgap ($E_g=$0.74~eV at 300~K) and high melting point ($\sim$1400~K) \cite{levinshtein1996handbook}. For In$_{0.53}$Ga$_{0.47}$As, $m_n^* = 0.041 m_e$ and $m_p^* = 0.45 m_e$, where $m_e$ is the electron mass. We also took into account the temperature-dependence of $E_g$ to adequately model the electric behavior of the cathode at high temperatures \cite{paul1991empirical,levinshtein1996handbook}.
The thickness of the cathode for this study is set to be 1.5~$\mu$m, which is thick enough to absorb the incident thermal radiation and is comparable to the electron diffusion length ($\sim3$~$\mu$m) in \textit{p}-doped In$_{0.53}$Ga$_{0.47}$As~\cite{ambree1992dependence}.
When the cathode absorbs thermal radiation above the bandgap energy, the electron distribution in the conduction band is perturbed from the equilibrium state due to photoexcitation. This perturbation can be described by adjusting the Fermi level for the quasi thermal equilibrium state.
The quasi-Fermi level for the conduction band can be written as~\cite{hatsopoulos1979thermionic},
\begin{equation}
E_{F,n}=E_F+k_B T_C \ln(n/n_{eq})
\label{eq:E_Fn}
\end{equation}
\noindent where $n$ is the electron concentration in the conduction band under photoexcitation, and 
$n_{eq} = N_C \exp\bigl[-(E_g-E_F)/k_B T_C\bigr]$ 
is the electron concentration in the conduction band at equilibrium. Here we assume that the energy level of the valance band is equal to zero ($E_{V}$=0). Equation (\ref{eq:E_Fn}) clearly indicates that the quasi-Fermi level is directly influenced by photoexcitation: the incoming photon energy increases the number of electrons in the conduction band of the cathode and, as a consequence, raises the quasi-Fermi level when the cathode temperature is at $T_C$.

While the calculation of $n_{\rm eq}$ is straightforward, the concentration of electrons in the conduction band upon photoexcitation, $n$, should be determined by considering the balance between the photoexcitation rate of electrons due to near-field radiation ($\dot{\Gamma}_{\rm NF}$), the recombination rate ($\dot{\Gamma}_{\rm R}$), and the net thermionic emission rate of electrons from the cathode to the anode ($\dot{\Gamma}_{\rm net}$):
\begin{equation}
\dot{\Gamma}_{\rm NF}-\dot{\Gamma}_{\rm R}-\dot{\Gamma}_{\rm net}=0
\label{Eq:Gamma}
\end{equation}
The near-field photoexcitation rate can be obtained from the formulation of near-field thermal radiation above the bandgap of the cathode, i.e., $\dot{\Gamma}_{\rm NF}=\left(1/d_c\right) \int_{\hbar\omega \geq E_g} \left\lbrace\left[S_z(T_E,\omega, d_g+d_c)-S_z(T_E,\omega,d_g)\right]/\hbar\omega \right\rbrace d\omega$, where $d_c$ is the cathode thickness and $\omega$ is the angular frequency. The near-field radiative heat flux at a certain point of the cathode can be obtained using the \textit{z}-component of the time-averaged Poynting vector that is formulated by fluctuational electrodynamics~\cite{park2008performance, francoeur2008role}:
\begin{equation}
S_z(T_E,\omega,z) = \frac{2k_0^2\Theta(\omega,T_E)}{\pi} \text{Re} \left\lbrace  i\varepsilon_E^{\prime\prime} \left[ G^{e}_{xj}G^{h*}_{yj} - G^{e}_{yj}G^{h*}_{xj}\right]_{j=x,y,z} \right\rbrace 
\label{Eq:NF}
\end{equation}
\begin{sloppypar}
\noindent where $k_0=\omega/c_0$ is the wavevector in vacuum, $\varepsilon_E^{\prime\prime}$ is the imaginary component of the thermal emitter's dielectric function, and $\Theta(\omega,T)=\hbar\omega/\left[ \exp\left( \hbar\omega/k_BT\right) -1\right]$ is the mean energy of the Planck oscillator. The electric dyadic Green's function, $G_{ij}^{e}({\bf x},{\bf x^\prime},\omega)$ in tensor notation, denotes the electric field at point ${\bf x}$ due to the current source at ${\bf x^\prime}$, and $G_{ij}^{h}$ is the tensor notation of the magnetic dyadic Green's function, which is defined as $\overline{\overline{\bf{G}^{\it h}}} = \nabla\times \overline{\overline{\bf{G}^{\it e}}}$ for non-magnetic materials. The detailed formulation of dyadic Green's functions for a multilayered structure can be found in previous works~\cite{francoeur2008role,park2008performance} and will not be repeated here. The superscript $*$ denotes the complex conjugate, and the subscript $j$ denotes the state of polarization of the source, which may involve a summation over the three orthogonal components. 
\end{sloppypar}
The recombination rate $\dot{\Gamma}_{\rm R}$ is determined by considering the near-field radiative recombination ($\dot{\Gamma}_{\rm R,NF}$), Auger recombination ($\dot{\Gamma}_{\rm Aug}$),  Shockley-Read-Hall (SRH) recombination ($\dot{\Gamma}_{\rm SRH}$), and surface recombination ($\dot{\Gamma}_{\rm Surf}$) mechanisms:
 \begin{equation}
 \dot{\Gamma}_{\rm R}=\dot{\Gamma}_{\rm R,NF}+\dot{\Gamma}_{\rm Aug}+\dot{\Gamma}_{\rm SRH}+\dot{\Gamma}_{\rm Surf}
 \end{equation}
The near-field radiative recombination rate is calculated by assuming that near-field thermal radiation from the cathode to the emitter is originated from the recombination of above-bandgap electrons in the cathode~\cite{shockley1961detailed,pla2007influence,francoeur2011thermal}.
The near-field radiative recombination rate, $\dot{\Gamma}_{\rm R,NF}$, can be written as 
\begin{equation}
\dot{\Gamma}_{\rm R,NF}=\frac{(1-C_{\rm PR})\left(np/n_{\rm eq}p_{\rm eq}-1\right)}{d_c}\int_{\hbar\omega \geq E_g} \frac{S_{-z}(T_C,\omega, z=0)}{\hbar\omega} d\omega
\label{eq:Gamma_R}
\end{equation}
\noindent where $S_{-z}(T_C,\omega,0)$ denotes the z-component of the time-averaged Poynting vector emitted from the cathode and received by thermal emitter interface at $z=0$. While $S_{-z}(T_C,\omega, 0)$ can be calculated using Eq. (4), $T_C$ should be used to calculate the mean energy of the Planck oscillator and a different form of the dyadic Green's function should be implemented to switch the source and receiver layers~\cite{francoeur2008role,park2008performance}. 
The coefficient $\left(np/n_{\rm eq}p_{\rm eq}-1\right)$, where $p$ is the hole concentration in the conduction band under photoexcitation and $p_{\rm eq} = N_V \exp\left(-E_F/k_BT_C\right)$ is the hole concentration at equilibrium, is multiplied to consider the increase of the electron-hole pair recombination rate due to the splitting of quasi-Fermi levels upon photoexcitation~\cite{wurfel2016physics}. 
In addition, a portion of photons emitted due to radiative recombination are re-absorbed within the material and regenerate electron-hole pairs. This photon recycling is also taken into account by defining the photon-recycling coefficient $C_{\rm PR}$. In the present work, we assumed $C_{\rm PR}$ to be 0.1 by following previous works~\cite{francoeur2011thermal,martin2004temperature,stollwerck2000characterization,letay2006simulating}.
Another electron recycling process may occur at the cathode interface when a portion of the emitted electrons are reflected back from the anode and reach the cathode conduction band, contributing to the increase of the thermionic power output and the efficiency~\cite{kribus2016solar}. However, the present study assumes that all the reflected electrons are lost due to surface recombination at the surface of the cathode.
The Auger recombination rate can be calculated from the following equation:
\begin{equation}
\dot{\Gamma}_{\rm Aug}=\dfrac{n-n_{\rm eq}}{\tau_{\rm Aug}}
\end{equation}
where $\tau_{\rm Aug}$ is the Auger lifetime. For a \textit{p}-type material under high injection, $\tau_{\rm Aug}$ is expressed as $\tau_{\rm Aug}=1/[(C_ n+C_ p)\times (n-n_{\rm eq})^2]$ \cite{kerr2002general},
where $C_n$ and $C_p$ are the Auger recombination coefficients for electrons and holes, respectively, and set to be $C_{n}=C_{p}=8.1\times10^{-29}$ cm$^6/$s  by following Ref.~\cite{ahrenkiel1998recombination}.
The SRH recombination rate is also estimated based on the assumption that defect-induced traps are located at the intrinsic energy level, which yields the following equation:
 \begin{equation}
 \dot{\Gamma}_{\rm SRH} = \dfrac{n \cdot p-n_{\rm eq}\cdot p_{\rm eq}}{\tau_n(p+n_i)+\tau_p(n+n_i)}
 \end{equation}
 where $\tau_{n(p)}$ is the SRH lifetime of electrons (holes). For the present study, both $\tau_{n}$ and $\tau_{p}$ were set to be 47.36 $\mu$m~\cite{ahrenkiel1998recombination}.
 Finally, the effect of surface recombination is considered by calculating the recombination rate occurring at the surface of the cathode,
  \begin{equation}
 \dot{\Gamma}_{\rm Surf}=qS(n-n_{\rm eq})/d_{\rm c}
 \label{Eq.Surf}
 \end{equation}
 where $q$ is the elementary charge and $S$ is the effective surface recombination velocity of electrons at the interface. For the present study, $S$ was assumed to be $10^4$ m/s~\cite{segev2013loss}. The contribution of each recombination rate to the overall recombination rate is discussed in details in the Supplementary Material: see Fig. S1.
 
The net thermionic emission rate of electrons can be calculated by  $\dot{\Gamma}_{\rm net}= \left(J_C - J_A\right)/{qd_c}$, where $J_{C(A)}$ is the current density emitted from the cathode (anode). The thermionic current density from the cathode is written as $J_C = q \int_{E_g+\chi_C}^\infty v_n g(E) f(E) dE$, where $\chi_{\rm C}$ is the electron affinity of the cathode, $v_n$ is the electron velocity perpendicular to the cathode surface, $g(E)$ is the electron density of states, and $f(E)$ is the Fermi-Dirac distribution~\cite{hatsopoulos1979thermionic}.
By assuming that the distribution of electrons above the electron affinity is approximated with a parabolic density of states and classical Maxwell-Boltzmann statistics, the thermionic current density of the cathode can be simplified as~\cite{schwede2010photon},
\begin{equation}
J_C = A^*T^{2}_{C} \exp\left[-\dfrac{\phi_C-(E_{F,n}-E_F)}{k_BT_C}\right]
\label{Eq:Jc}
\end{equation}
where $A^* = 4\pi qm^*_n k_B^2/h^3$ is the material-specific Richardson-Dushman constant~\cite{musho2013quantum}, and  $\phi_C$ is the work function of the cathode. The derivation of Eq. (\ref{Eq:Jc}) is described in the Supplementary Material. As illustrated in Fig. \ref{Fig.1}(b), the work function can be correlated with the electron affinity $\chi$ as $\phi=\chi + E_g - E_F$.  
Equation (\ref{Eq:Jc}) suggests that near-field illumination on the ${p}$-doped semiconductor lowers the energy barrier for electron emission by the difference between the quasi-Fermi level and the Fermi level at equilibrium, resulting in an exponential increase in cathode current density. 
If the cathode is not exposed to near-field radiation, Eq. (\ref{Eq:Jc}) is reduced to the standard thermionic current density equation, i.e., $J_C = A^*T^{2}_{C} \exp\left({-\phi_C}/{k_BT_C}\right)$. 
The same equation can be used to describe the thermionic current density from the anode by $J_A =A^*T_A^{2} \exp\left({-\phi_A}/{k_BT_A}\right)$, where $T_A$ and $\phi_A$ are the temperature and the work function of the anode, respectively, and  $A^*$ should be determined based on the anode material. 

Along with the concentration of electrons in the conduction band of the cathode under the illumination of near-field thermal radiation, another key parameter in the performance analysis of the NETEC system is the cathode temperature. To this end, the energy balance within the cathode should be carefully analyzed. Under the assumption that the cathode has a uniform temperature due to its small thickness (i.e., 1.5~$\mu$m), the energy balance of the cathode can be written as
\begin{equation}
\dot{Q}_{E\rightarrow C}=\dot{Q}_{C \rightarrow E}+\dot{Q}_{C\rightarrow A}+(J_C-J_A)\phi_C + 2\dfrac{k_B}{q}(J_C T_C - J_A T_A)
\label{Eq:T}
\end{equation}
Here, $\dot{Q}_{i \rightarrow j}$ denotes the radiative heat flux emitted from $i$ and absorbed in $j$, where subscripts $i$ and $j$ can be $E$, $C$, and $A$ indicating the emitter, cathode, and anode, respectively.  For example, $\dot{Q}_{E \rightarrow C}$ is the near-field radiative heat flux emitted from the thermal emitter and absorbed by the cathode, which can be calculated by 
$\dot{Q}_{E \rightarrow C}=\int_{\hbar\omega \geq E_g} \left[S_z(T_E,\omega, d_g+d_c)-S_z(T_E,\omega,d_g)\right] d\omega$. Similarly, $\dot{Q}_{C \rightarrow E}$ can be calculated by $\dot{Q}_{C \rightarrow E}=\int_{\hbar\omega \geq E_g} S_{-z}(T_C,\omega, 0)d\omega$ based on the reciprocity, as described in Eq. (\ref{eq:Gamma_R}), and $\dot{Q}_{C \rightarrow A}$ the same way by considering the cathode as a source layer and the anode as an absorbing medium. However, $\dot{Q}_{A \rightarrow C}$ is ignored due to the relatively small thermal radiation from the anode at its low temperature. $(J_C-J_A)\phi_C$ indicates the electric energy that is taken from the cathode due to thermionic emission. Finally, the kinetic energy carried away by the electrons emitted from either the cathode or anode is taken into account by the last term~\citep{hatsopoulos1979thermionic,segev2012efficiency}. By solving Eqs. (\ref{Eq:Gamma}) and (\ref{Eq:T}) simultaneously, we can obtain the cathode temperature ($T_{\rm C}$) and the total electron concentration in the conduction band ($n$). Consequently, the current density of the cathode can be calculated from Eq. (\ref{Eq:Jc}).

\section{Results and Discussion}

\begin{figure}[t!]
\centering
\includegraphics[width=0.65\linewidth]{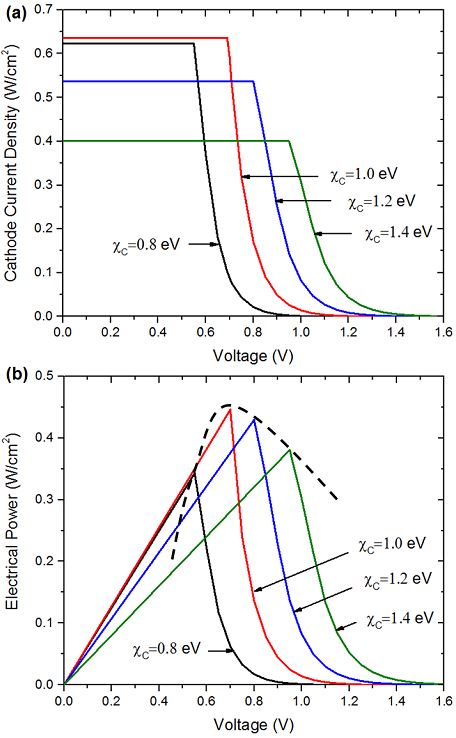}
\caption{(a) The ideal $J$-$V$ characteristics of the cathode and (b) the electrical power density as a function of the operating voltage for different electron affinities of the cathode when $T_E=1200$~K and $T_A=300$~K. The anode work function was set to 0.7~eV, and the emitter-cathode gap distance was assumed to be $d_g=100$~nm.
}
\label{Fig.2}
\end{figure}

Figure \ref{Fig.2} shows the ideal current-voltage characteristics and the corresponding power generation of the NETEC system for different electron affinities of the cathode (In$_{0.53}$Ga$_{0.47}$As) when tungsten is used as a broadband thermal emitter. 
For the calculation, the emitter temperature ($T_E$) was set to be 1200~K, while the anode temperature ($T_A$) was 300~K. The vacuum gap distance between the emitter and the cathode was 100nm, and the work function of the anode ($\phi_A$) was set to 0.7~eV.
As shown in Fig.\ref{Fig.2}(a), the $J$-$V$ characteristic curves of the NETEC system are similar to those of standard thermionic emission with two distinctive regimes~\cite{hatsopoulos1979thermionic}. When the operating voltage is smaller than the difference between the cathode and anode work functions (i.e., $qV_{\rm op} \leq \Delta \phi$, where $\Delta\phi = \phi_C - \phi_A$), the cathode current density is fully saturated to have a flat band. However, as the operating voltage further increases above the flat-band condition (i.e., $qV_{\rm op}>\Delta \phi$), the current density exponentially decays because $qV_{\rm op}$ increases the energy barrier against electron emission from the cathode: see Fig. \ref{Fig.1}(b). This characteristic can be better understood by rewriting Eq.(\ref{Eq:Jc}) in the following form~\cite{hatsopoulos1979thermionic,segev2012efficiency}:
\begin{equation}
\begin{aligned}
J_C &= qn\left\langle v_n \right\rangle \exp\left(-\frac{\chi_C}{k_BT_C}\right)~ &{\rm for}~ qV_{\rm op}\leq \Delta \phi \\
J_C &= qn\left\langle v_n \right\rangle \exp\left(-\frac{\chi_C+qV_{\rm op}-\Delta \phi}{k_BT_C}\right) ~ &{\rm for}~qV_{\rm op} > \Delta \phi 
\label{eq:JcMod}
\end{aligned}
\end{equation}
where $\left\langle v_n \right\rangle = \left(k_B T_C/2\pi m_e^*\right)^{1/2}$ is the average electron velocity perpendicular to the surface. 
Equation (\ref{eq:JcMod}) suggests that the near-field enhanced photoexcitation of electrons should have a direct impact to the thermionic current density by increasing electron concentration in the conduction band of the cathode. In addition, the current density increases as the thermal energy of the cathode ($k_B T_C$) increases or the electron affinity of the cathode ($\chi_{\rm C}$) decreases. This inverse relation between $\chi_{\rm C}$ and $J_C$ can be confirmed in Fig.\ref{Fig.2}(a) for  $\chi_{\rm C}$ bigger than 1.0~eV. However, the flat-band current density begins to decrease when $\chi_{\rm C}$ is below 0.8~eV. We believe that this unique feature is the result of the competing effect of $\chi_{\rm C}$ to the cathode current density and the cathode temperature: while lowering the electron affinity increases the electron emission rate from the cathode, too low electron affinity may reduce the temperature of the cathode as a result of energy balance, which adversely affects the current density. Moreover, low cathode electron affinities may also increase the reverse current flow from the anode, which reduces the net current density of the NETEC system. More details about the effect of the electron affinity to the cathode temperature will be discussed in Fig. \ref{Fig.4}(c). 

The power density of the NETEC system can be calculated using $P_{\rm NETEC}=(J_C-J_A)\Delta \phi/q$, which is in fact the maximum power that the NETEC system can produce at $qV_{\rm op} = \Delta \phi$ for a given electron affinity of the cathode \cite{hatsopoulos1979thermionic}.
Figure~\ref{Fig.2}(b) shows the change of $P_{\rm NETEC}$ for different electron affinities, demonstrating that there should be an optimum cathode electron affinity that maximizes $P_{\rm NETEC}$ for a given operational condition: for example, $P_{\rm NETEC}$ in Fig.~\ref{Fig.2}(b) becomes maximum at 0.45~${\rm W/cm^2}$ at ${\rm \chi_C}=0.96$~eV, which corresponds to $V_{\rm op} = 0.68{\rm V}$.
Previous studies have demonstrated that the work function, or more specifically the electron affinity of a material can be manipulated by applying surface coatings~\cite{de2005tuning}. Alkali or alkali-earth metals, mostly cesium (Cs), have been widely used to lower the work function of a material. For example, cesiated tungsten has a much lower work function ($\sim$1.7~eV) than that of pure tungsten ($\sim$4.5~eV)~\cite{fomenko1966handbook}, and even can be lowered to $\sim$1~eV by interacting cesium and oxygen on tungsten~\cite{desplat1980interaction}.
Besides tungsten, other materials also have been examined in efforts to achieve a low work function. Yi \textit{et al}.\cite{yi2011effective} found that depositing a subnanometer Al layer onto a multilayered graphene film could reduce the work function from 4.40~eV to $\sim$3.77~eV. 
Several research groups demonstrated that the electron affinity of semiconductors, such as GaAs and Si, can be manipulated over a wide range, even producing negative affinities, with cesium deposition \cite{huang2007low,madey1971electron,martinelli1974thermionic,smestad2004conversion,schwede2010photon}.
Cesium coating is not the only way to control the electron affinity. It has been shown that co-adsorption of oxygen and potassium on a Si surface can also lower the work function by 0.7~eV \cite{morikawa1995further}.

\begin{figure}[t!]
\centering
\includegraphics[width=0.75\linewidth]{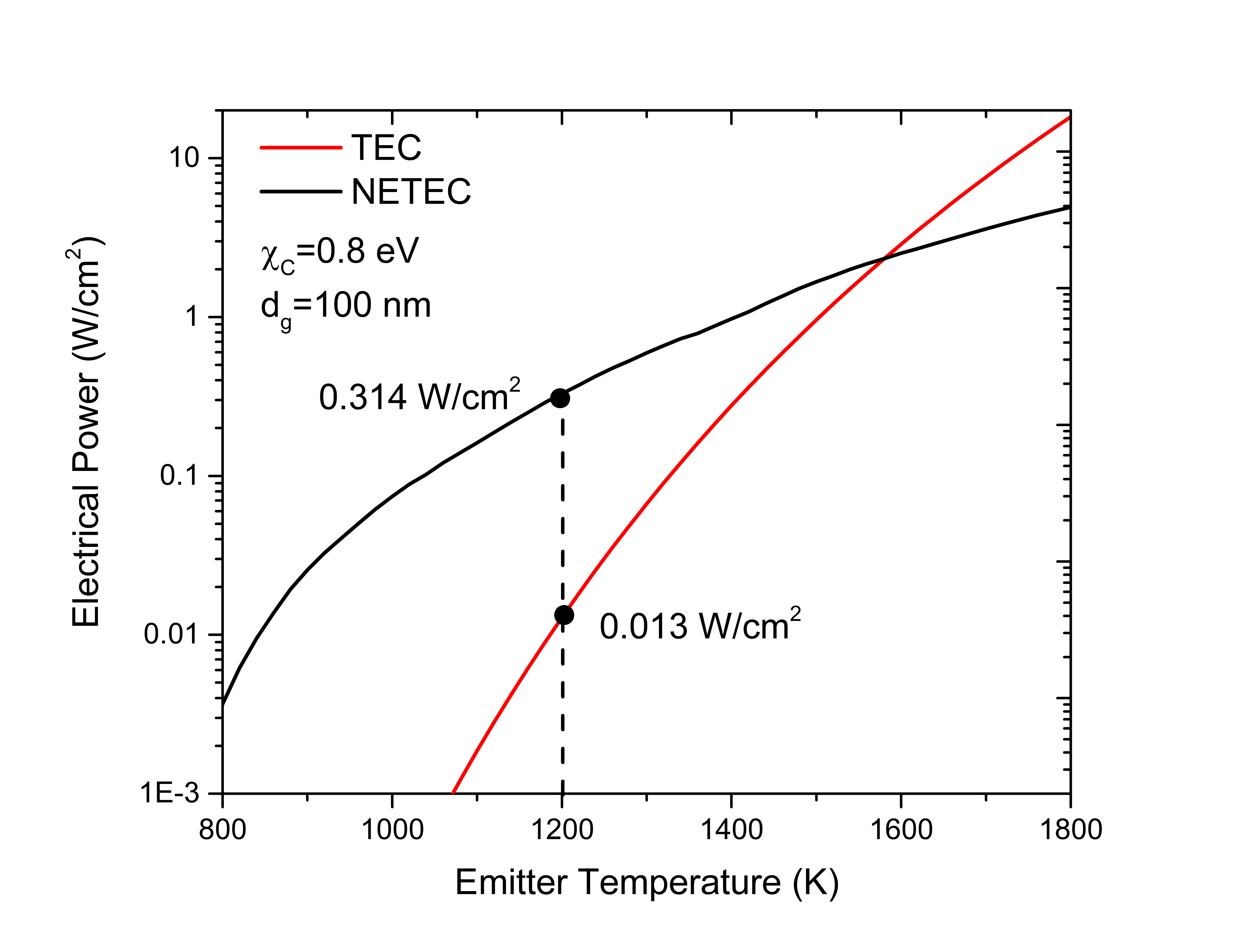}
\caption{ Comparison of power generation contributions of conventional thermionic (TEC) and near-field enhanced thermionic (NETEC) energy conversions as a function of the emitter temperature. Here, $T_A=300$~K and $\phi_A=0.7$~eV, and the emitter-cathode gap distance was assumed to be $d_g=100$~nm.
}
\label{Fig.3}
\end{figure}

\begin{figure}[t!]
\centering
\includegraphics[width=\linewidth]{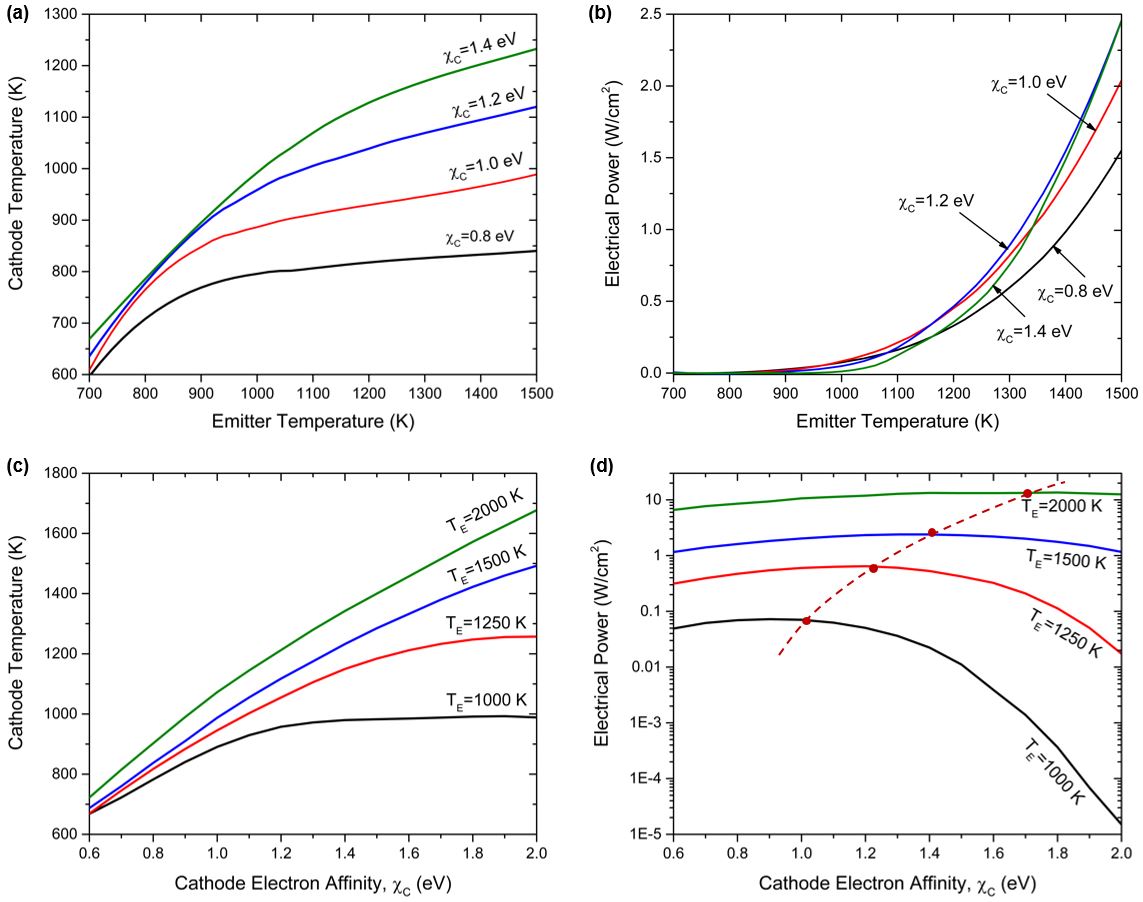}
\caption{The temperature of the cathode calculated from the energy balance and it's corresponding power output as a function of emitter temperature [(a) and (b)] and the cathode electron affinity [(c) and (d)]. For these simulations, the temperature of the anode and its work function were set to 300~K and 0.7~eV, respectively. The vacuum gap distance between emitter and cathode is assumed to be 100~nm.}
\label{Fig.4}
\end{figure}

Figure \ref{Fig.3} demonstrates the near-field enhancement of thermionic power generation by comparing the power densities of NETEC and conventional thermionic energy conversion (TEC) systems operating at the same heat source and anode temperatures. For computation, we assume that both systems have a In$_{0.53}$Ga$_{0.47}$As cathode with $\chi_{\rm C}=0.8$~eV and a tungsten anode with $\phi_A = 0.7$~eV maintained at 300~K. However, the hot sides of the two systems have different configurations: the NETEC system has a tungsten thermal emitter to allow near-field thermal radiation to the cathode across a 100-nm gap while the TEC system has the cathode directly attached to the source to allow thermionic emission only. 
At low source temperatures below 800~K, only a small portion of electrons would be able to overcome the work function of the cathode due to low electron velocities obtained from cathode's thermal energy. The near-field enhancement of thermionic emission becomes prominent in the heat source temperature range of 800~K~$<T_E<$~1600~K. In this range, photoexcited electrons are redistributed through thermalization to have orders of magnitude more electrons above the electron affinity of the cathode. For example, when the source temperature is 1200~K, the power generation obtained from the NETEC system ($P_{\rm NETEC} = 0.314$~W/cm$^2$) exceeds that of the TEC system ($P_{\rm TEC} = 0.013$~W/cm$^2$) operating at the same source temperature by more than 20 times. However, the TEC system generates more electric power than the NETEC as the source temperature further increases over $>\sim$1600~K. This trend can be better understood by observing the cathode temperature of the NETEC system as a function of the source (or thermal emitter) temperature. As shown in Fig. \ref{Fig.4}(a) for $\chi_{\rm C}$=0.8~eV, the cathode temperature of the NETEC system is almost flat at around 800~K while the source temperature increases to 1500~K, indicating that the most of near-field thermal radiation increased by higher emitter temperatures is used to photoexcite more electrons instead of heating the cathode. Therefore, the increase of the NETEC power generation at higher source temperatures is mainly owing to the increase of photoexcited electrons rather than the increase of thermal energy, which ultimately leads to less electrical power generation than when using the heat source purely for thermionic power generation. This observation reveals that the NETEC power generation can lower a source temperature required for thermionic energy conversion. 

We also compared the NETEC system with the near-field thermophotovoltaic (NTPV) system that was configured with a tungsten thermal emitter separated from a In$_{\rm 0.53}$Ga$_{\rm 0.47}$As TPV cell by a 100$~$nm gap distance, following the configurations used in Refs.  \cite{park2008performance,francoeur2011thermal}. The thermal emitter temperature is 1500~K and the backside of the TPV cell is exposed to a cooling fluid at 300 K with the convection heat transfer coefficient of 10$^3$~W/m$^2$-K~\cite{incropera2002fundamentals}. By considering surface recombination as the only recombination mechanism, the NTPV system generates electrical power of $0.92$~W/cm$^2$, which is about 1.7 times smaller than that of the NETEC system ($P_{\rm NETEC}=1.57$~W/cm$^2$) under the same thermodynamic conditions. Although a higher power could be generated in the NTPV system by further cooling the TPV cell with a higher convection heat transfer coefficient (e.g., 2.1~W/cm$^2$ when using 10$^4$~W/m$^2$-K), it may need a phase change cooling scheme that would add more complexities to the system operation. 

To better understand the performance of the NETEC system, we have calculated the cathode temperature and the power density as a function of emitter temperature and cathode electron affinity. For this calculation, the emitter-cathode gap distance is set to be 100 nm, and the work function of the anode is 0.7 eV.
Figure \ref{Fig.4}(a) demonstrates that the the cathode temperature increases with the increasing emitter temperature, but with a much smaller slope, i.e., $\partial T_C/\partial T_E$, at higher emitter temperatures. 
As discussed in Fig. \ref{Fig.3}, this trend indicates that the  increase of near-field thermal radiation at higher emitter temperatures is dominantly used to photoexcite electrons to the conduction band of the cathode. Therefore, the increase of NETEC power generation shown in Fig. \ref{Fig.4}(b) is owing to the contribution of near-field enhanced photoexcitation rather than the thermal effect. 
However, the effect of the emitter temperature onto the NETEC power output is complicated, depending on the the electron affinity of the cathode, $\chi_C$. 
The effect of $\chi_C$ on the cathode temperature and the NETEC power generation is more clearly shown in Figs. \ref{Fig.4}(c) and (d). Since the electron affinity represents the binding energy of electrons at the semiconductor-vacuum interface, the increasing electron affinity of the cathode raises the cathode temperature, ultimately making it very close to the emitter temperature. Once the cathode is in thermal equilibrium with the emitter, there should be no near-field thermal radiation between them.
As a result, the NETEC power generation rapidly decreases as $\chi_C$ increases particularly at low emitter temperatures: see Fig. \ref{Fig.4}(d). In this  $\chi_C$ range, power generation is solely due to thermionic (or thermal) contribution. 
While not shown clearly in the log scale, it should be noted that there is an optimal $\chi_C$ value maximizing the NETEC power at each emitter temperature: see the dashed curve in the figure. As discussed in Fig.\ref{Fig.2}(b), although a small electron affinity of the cathode is desired to promote the cathode current density, too small affinities decreases the cathode temperature while increasing the inverse current flow, both of which adversely affect NETEC power generation. 

\begin{figure}
\begin{center}
    \includegraphics[width=0.65\linewidth]{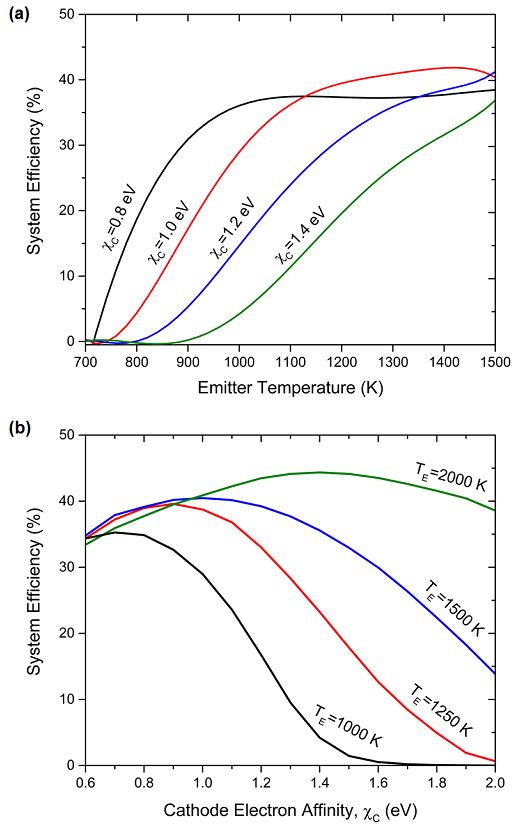}
\end{center}  
\caption{(a) The efficiency of the NETEC system as a function of (a) the emitter temperature, and (b) the cathode electron affinity. Under the same conditions as Fig. 4.} 
\label{Fig.5}
\end{figure}

As mentioned in the introduction, the ideal energy conversion efficiency of photon-enhanced thermionic emission has been predicted to be above 50\%~\cite{schwede2010photon}, and even when loss mechanisms are considered, the efficiency is still in the range of 30-40\%~\cite{segev2012efficiency,segev2013loss,su2013parametric}.
For the case of NETEC system, its energy conversion efficiency should be defined as the ratio of the electrical power output, $P_{\rm NETEC}$, to the heat input to the NETEC system, or near-field thermal radiation absorbed by the cathode, $Q_{E\rightarrow C}$:
\begin{equation}
\eta_{\rm NETEC}=\dfrac{P_{\rm NETEC}}{Q_{E \rightarrow C}}
\label{Eq:Eff}
\end{equation}
Figure \ref{Fig.5} shows the efficiency of the NETEC system for different emitter temperatures, $T_{E}$, and cathode electron affinities, $\chi_{\rm C}$, respectively. 
The predicted NETEC efficiency is in between 30-40\% for practical $T_{E}$ and $\chi_{\rm C}$ ranges, which is similar to the aforementioned efficiency range of photon-enhanced thermionic emission and much higher than that of near-field thermophotovoltaics. The main reason of such high efficiency for the NETEC system can be found from the effective use of excessive photon energy through thermalization. 
At $\chi_{\rm C}=0.8$~eV in Fig. \ref{Fig.5}(a), the system efficiency rapidly increases with the increasing emitter temperature (and consequently the cathode temperature)  up to 1000~K. In this region, near-field thermal radiation directly impact the photoexcitation of electrons to generate more electric power. 
However, further increase of the emitter temperature above 1000 K does not increase but rather slightly decreases the efficiency.
As the emitter temperature increases, the work function of the cathode (i.e., $\phi_{C}=\chi_{C}+E_{g}-E_{F}$) becomes smaller due to the adverse temperature dependence of the energy bandgap \cite{hatsopoulos1979thermionic}. The decrease of $\Delta\phi$ slows down the increasing rate of $P_{\rm NETEC}$ as the emitter temperature keeps increasing, which ultimately lowers the efficiency. Similar patterns are observed for higher $\chi_{C}$ values with the shift of curves towards higher emitter temperatures. 
In Fig. \ref{Fig.5}(b), the system efficiency also shows a increasing and decreasing pattern as the cathode electron affinity increases
We believe that this behavior in general follows that of the NETEC power generation shown in Fig. \ref{Fig.4}(d). However, it should be noted that the optimal $\chi_C$ values for the best system efficiency are not identical to those for the maximum power generation. This deviation comes from the effect of the cathode electron affinity onto $Q_{E\rightarrow C}$. At the same emitter temperature, higher electron affinities causes more thermalization-induced heating of the cathode, which narrows the energy bandgap of the cathode to increase more absorption of near-field thermal radiation (i.e., $Q_{E\rightarrow C}$) in turn. 
As a result, the efficiency becomes maximum at a slightly lower $\chi_C$ than power output, and more rapidly decreases as $\chi_C$ further increases. 
This comparison suggests that the electron affinity of the cathode should be carefully determined based on the temperature of a heat source, and more attention should be paid to the system efficiency. 

\begin{figure}
\begin{center}
    \includegraphics[width=0.75\linewidth]{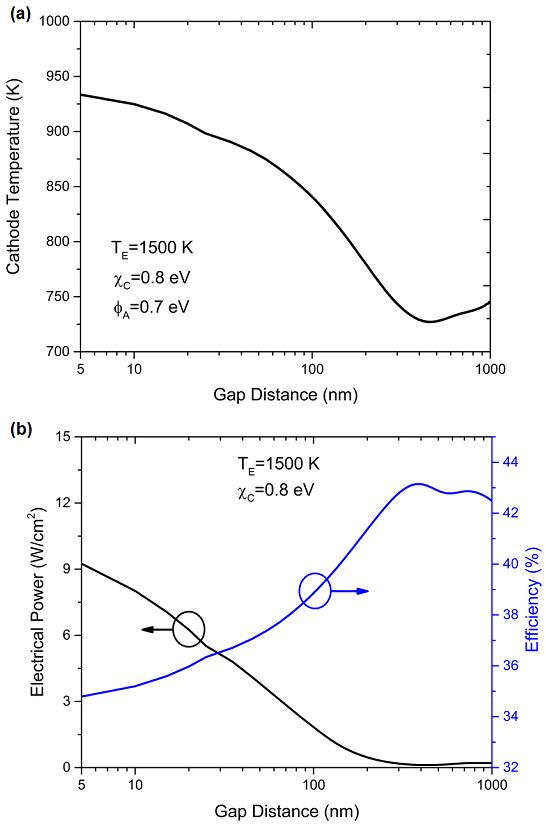}
\end{center}  
\caption{Effect of vacuum gap distance between the emitter and cathode on (a) the temperature of the cathode, and (b) the electrical power output and efficiency of the NETEC system. For this calculation, $T_E$ and $T_A$ are 1500~K and 300~K, respectively, with $\chi_{C}=0.8$~eV and $\phi_A=0.7$~eV.} 
\label{Fig.6}
\end{figure}

 The near-field effect on the NETEC is well described in Fig. \ref{Fig.6} that shows the emitter-cathode gap dependence of its power output and efficiency. For calculation, we assumed $T_{E}=1500$~K for a tungsten thermal emitter, and the electron affinity of the cathode and the work function of the anode were set to  $\chi_{\rm C}=0.8$~eV and $\phi_A=0.7$~eV, respectively. 
The overall gap-dependence of the NETEC power output follows a similar trend as near-field thermal radiation\cite{park2008performance}. As the gap distance decreases, net radiative heat transfer from the emitter to the cathode increases by several orders of magnitude, leading to a significant enhancement in the cathode temperature and the NETEC power generation. 
As the gap decreases from 500nm to 10~nm, for example, power generation increases from 0.12~W/cm$^2$ to 7.34~W/cm$^2$ to achieve more than 61 fold enhancement. However, when the gap is above 500nm, the near-field effect phases out with fringes due to the interference of far-field electromagnetic waves in the emitter-cathode vacuum gap.
The interference effect is more prominent in the efficiency curve when the gap distance is larger than 400nm. For the gap distance below 400nm, the efficiency decreases monotonically from $\sim$43\% at $d_g$=400~nm to $\sim$35\% at $d_g$=10~nm. 
In such small gap distances, the penetration depth of near-field thermal radiation is restricted to near the top surface of the cathode \cite{park2008performance}. Photoexcited electrons have more probabilities to be recombined before they diffusively travel to the bottom surface of the cathode. A thinner cathode may resolve the efficiency degradation and even further increase the NETEC power output. Nevertheless, it should be noted that the lowest NETEC efficiency is still $\sim$34\%, which is much higher than other solid-state energy conversion techniques. 

Although the proposed NETEC presents a novel way to make use of near-field thermal radiation for energy conversion with a unprecedentedly high energy conversion efficiency and enhanced power output, it should be noted that our model has been developed based on ideal assumptions to demonstrate the feasibility of the proposed concept. We have not considered potential challenges in realizing the NETEC system, such as the buildup of space charges between the cathode and anode and the lack of understanding on the thermal stability of cesiated low-bandgap semiconductors at high temperatures.
Space charges between electrodes can decrease the output power and efficiency of the thermionic energy conversion due to the repelling between the electrons traversing the gap between the cathode and anode \cite{lee2012optimal,meir2013highly,moyzhes2005thermionic}. To mitigate the space charge buildup issue in thermionic systems, previous studies have proposed a triode configuration, instead of a simple diode configuration, in which space charge clouds can be diminished by inserting a positively charged gate electrode in the cathode-anode space and applying longitudinal magnetic fields across the electrodes \cite{meir2013highly,moyzhes2005thermionic}. 
We believe that the same scheme can be adopted to the NETEC system with slight modification to resolve the space charge buildup.
Another challenge about the NETEC system is the lack of understanding on the thermal stability of low-bandgap semiconductors at high temperatures. Although we have chosen In$_{0.53}$Ga$_{0.47}$As as a cathode material for this study due to its low energy bandgap (0.74~eV) and high melting temperature ($\sim$1400~K), high-temperature behaviors of this material and III-IV compounds in general still remain in question. Moreover, cesium-coated surfaces may become unstable at high temperatures by evaporating excessive cesium ions (Cs$^+$), leading to the unstable electron affinity of the cathode \cite{uebbing1970behavior,desplat1980interaction,papageorgopoulos1981thermal}. 
Several groups have investigated the thermophysical properties of InGaAs alloys for various compositions up to the temperature of $\sim$1500~K and found no phase change or peculiar material behaviors at high temperatures. However, there is still lack of solid discussion regarding the high-temperature stability and change of electrical properties of InGaAs \cite{stillman1976electrical,touloukian1974thermophysical,hall1963solubility}. The spatial distribution of photo-generated electrons and recombination processes, the effect of junctions and realistic contacts also need to be considered in a future work for a more comprehensive understanding of the NETEC system.
Despite these challenges, however, we strongly believe that the NETEC opens a new direction of using near-field thermal radiation for renewable energy harvesting and recycling that may overcome the current challenges of near-field thermophotovoltaics.

\section{Conclusion}
In this article we have proposed a hybrid energy conversion system that combines near-field thermal radiation with thermionic emission. The proposed near-field enhanced thermionic energy conversion (NETEC) system makes use of near-field enhanced photoexcitation as well as thermal excitation of electrons in a low-bandgap semiconductor cathode to enhance the thermionic current density. 
We have demonstrated the remarkable enhancement in power output and efficiency achievable by the NETEC system. Moreover, near-field enhanced photoexcitation allows the operation of the NETEC system at much lower temperatures than the conventional thermionic energy conversion system. 
From the obtained results presented here, the concept of harvesting near-field thermal radiation through combined photovoltaic/thermal processes can provide a highly attractive way of recycling waste heat. 

\section{Acknowledgements}
This work has been supported by the National Science Foundation (NSF CBET-1403072 and ECCS-1611320). The authors would like to thank Prof. Mathieu Francoeur at the University of Utah for discussions and his valuable comments. 

\clearpage
\section*{References}
\bibliographystyle{elsarticle-num} 
\bibliography{References_EES}

\clearpage

\clearpage
\clearpage
\clearpage

\clearpage

\clearpage

\end{document}